\documentclass[12pt]{iopart}
\usepackage{iopams}  
\usepackage{graphicx}
\usepackage{color}
\usepackage{hyperref}
\usepackage{tikz} 
\usepackage{url}
\usepackage{aas_macros}
\usepackage[T1]{fontenc}
\newcommand{\Rs}{R_{\mathrm{S}}}
\newcommand{\com}{\mathcal{C}}
\newcommand{\rhoc}{\epsilon_{\mathrm{c}}}
\newcommand{\cs}{c_{\mathrm{s}}}
\newcommand{\stab}{\beta_{\mathrm{stab}}}

\eqnobysec
\begin{document}
%%%%%%%%%%%%%%%%%%% TITLE PAGE %%%%%%%%%%%%%%%%%%%%%%%%%%

\title[Slowly rotating Tolman VII solution]{Slowly rotating Tolman VII solution}

\author{Camilo Posada and Zden\v{e}k Stuchl\'ik}%\email{camilo.posada@physics.slu.cz}
%}\email{zdenek.stuchlik@physics.slu.cz}
\address{Research Centre for Theoretical Physics and Astrophysics, Institute of Physics, Silesian University in Opava, Bezru\v{c}ovo n\'{a}m. 13, CZ-746 01 Opava, Czech Republic}
\ead{camilo.posada@physics.slu.cz and zdenek.stuchlik@physics.slu.cz}

\begin{abstract}

We present a model of a slowly rotating Tolman VII (T-VII) fluid sphere, at second order in the angular velocity. The structure of this configuration is obtained by integrating numerically the Hartle-Thorne equations for slowly rotating relativistic masses. We consider a sequence of models where we vary the parameter $R/\Rs$, where $R$ is the radius of the configuration and $\Rs$ is its Schwarzschild radius, representing an adiabatic and quasi-stationary contraction by progressively reducing the radius while keeping the angular momentum and gravitational mass constant. We determined the moment of inertia $I$, mass quadrupole moment $Q$, and the ellipticity $\varepsilon$, for various configurations. Similarly to previous results for Maclaurin and polytropic spheroids, in slow rotation, we found a change in the behaviour of the ellipticity when $R/\Rs$ reaches a certain critical value. Based on our analysis for the T-VII solution, we found variations of $\mathcal{O}(10\%)$ in the $I-\mathcal{C}$ and $Q-\mathcal{C}$ relations, and $\mathcal{O}(1\%)$ variation in the $I-Q$ relation, with respect to the universal fittings proposed for realistic neutron stars. Our results suggest that the T-VII solution can be considered a rather good approximation for the description of the interior of neutron stars. 

\end{abstract}


%%%%%%%%%%%%%%%%%%%%%%%%%%%%%%%%%%%%%%%%%%%%%%%%%%

\section{Introduction}

The study of the astrophysics of neutron stars (NSs) may provide key information, not only about nuclear matter at extremely high densities (above nuclear saturation density) but also about the physics in the so-called strong gravity regime. A crucial element in this study is the equation of state (EOS), meaning the functional relation between pressure and mass-energy density of matter, which is commonly assumed to be described by a perfect fluid. Although various reasonable tabulated EOSs for realistic NSs have been proposed in the literature (see e.g.~\cite{Schaffner-Bielich:2020psc}), a definitive relation that can describe satisfactorily the internal constitution of NSs remains one of the biggest puzzles in relativistic astrophysics.  

Finding solutions to Einstein's equations, given certain EOS, is useful to connect observables like mass, radii, Love numbers, etc., with the NS internal structure. For instance, one alternative is to study configurations with polytropic EOS \cite{Stuchlik:2016xiq,Novotny:2017cep,Posada:2020svn} and simulate the realistic EOS by sequences of polytropic configurations \cite{Yagi:2014qua,Alvarez-Castillo:2017qki}. However, most of these solutions must be computed numerically due to the complexity of the field equations. Another alternative is to find analytic solutions to Einstein's equations that can describe, approximately, the interior structure of NSs. In this context, Schwarschild's interior solution for an incompressible fluid~\cite{Schwarzschild:1916inc,Stuchlik:2000ud,Boehmer:2003uz}, the Buchdahl model~\cite{Buchdahl:1967ApJ}, and the Tolman VII (T-VII) solution~\cite{Tolman:1939}, have been commonly employed in the study of the interior properties of NSs \cite{Lattimer:2000nx,Postnikov:2010yn}, and also in the context of the so-called $I$-Love-$Q$ relations \cite{Yagi:2013awa,Yagi:2016ejg}. The T-VII solution is characterized by possessing an energy density that varies quadratically with the radial coordinate; this behaviour turns out to be eminently reasonable for the intermediate regions of realistic NSs \cite{Lattimer:2000nx}, and this is why the T-VII model has attracted considerable interest in the literature \cite{Neary:2001ai,Tsui:2005,Tsui:2005hb,Raghoonundun:2015wga,Moustakidis:2016ndw,Sotani:2018aiz,Jiang:2019vmf,Jiang:2020uvb,Stuchlik:2021coc,Posada:2021zxk,Stuchlik:2021vdd,Yagi:2021st,Posada:2022lij, Pappas:2022gtt, Koliogiannis:2022uim, Jayawiguna:2022ftj}.

Recent measurements carried out by the Neutron Star Interior Composition Explorer (NICER)~\cite{nicer}, have placed important constraints on values of masses and radii of rotation-powered X-ray pulsars~\cite{Miller:2021qha, Riley:2021pdl,Salmi:2022cgy}. Thus, the study of the rotational properties of compact objects is of great interest, and numerous approaches have been proposed in the context of GR \cite{Paschalidis:2016vmz}. One of them is the perturbative method for slowly rotating relativistic masses, up to second order in the rotational speed $\Omega$, developed by Hartle and Thorne (HT)~\cite{Hartle:1967he,Hartle:1968si}. In this approach, a static and spherically symmetric background configuration with mass $M$ and radius $R$ is set into slow uniform rotation, such that, $\delta=\Omega/\Omega_{0}<<1$, where, $\Omega_{0}=\sqrt{GM/4\pi^2R^3}$ is the mass-shedding frequency. Thus, fractional changes in pressure, energy density, and gravitational field are much less than unity. The slowly rotating approximation is reasonable for most of the known pulsars with $\Omega<300~\mathrm{Hz}$, for instance, PSR J$0030+0451$, whose frequency was estimated to be $f=205.53~\mathrm{Hz}$, so $\delta=0.14$ \cite{Silva:2020acr}.  Even for PSR J$0740+6620$, which spins at $346.53~\mathrm{Hz}$ \cite{Yunes:2022ldq}, $\delta\sim 0.16$ thus validating the slow rotation approximation. 

The HT formalism has been applied to various tabulated EOSs for NSs, since their pioneering work \cite{Hartle:1968si} and more recently by \cite{Silva:2020acr}. In the context of solutions to Einstein's equations for compact objects, the HT perturbative method has been applied to uniform density configurations \cite{Chandra:1974, Petroff:2007tz, Reina:2015jia} and polytropic fluid spheres \cite{Miller:1977, Benitez:2020fup}. Slowly rotating T-VII models, to the first order in $\Omega$, were considered by \cite{Jiang:2020uvb,Stuchlik:2021vdd}. By applying a post-Minkowskian perturbation method, \cite{Jiang:2020uvb} found analytical solutions for the ‘dragging' function $\omega$ and moment of inertia $I$. However, a full treatment to the second order in $\Omega$ is still missing in the literature. Given the complexity of the HT structure equations at order $\Omega^2$, as well as the T-VII solution, it is not possible to obtain the corresponding rotating solution, within this approximation, in analytical form. Thus, in this paper, we present a slowly rotating T-VII solution, by numerically solving the HT structure equations at order $\Omega^2$.

The organization of the paper is as follows. In~\sref{sec:Hartle} we review the perturbative HT method for slowly rotating relativistic masses, at the second order in the angular frequency $\Omega$. In~\sref{sec:Tolman} we discuss briefly the T-VII solution. In~\sref{sec:Results} we present our numerical results for the integral and surface properties of slowly rotating T-VII fluid spheres. Finally, ~\sref{sec:Conclud} contains our summary and conclusions.    

\emph{Conventions and notation:} We use geometrized units $(c=G=1)$, unless stated otherwise, and we adopt the signature $(-,+,+,+)$ for the metric.

%%%%%%%%%%%%%%%%%%%%%%%%%%%%%%%

\section{Hartle-Thorne framework for slowly rotating relativistic stars}\label{sec:Hartle}

In this section, we review the HT method for slowly rotating relativistic masses \cite{Hartle:1967he,Hartle:1968si}. We follow the notation and conventions used in \cite{Hartle:1968si, Chandra:1974, Miller:1977}. 

\subsection{Non-rotating configuration in hydrostatic equilibrium}

The starting point is a non-rotating configuration which is described by a metric in the spherically symmetric Schwarzschild-like form 
 \begin{equation}\label{statmetric}
\rmd s^2 = -\rme^{\nu(r)}\rmd t^2+\rme^{\lambda(r)}dr^2 + r^2 (\rmd\theta^2 + \sin^2\theta \rmd\phi^2),
\end{equation}
where 
\begin{equation}\label{misner-sharp}
\rme^{-\lambda(r)}\equiv 1-\frac{2m(r)}{r},\quad m(r)=4\pi\int_{0}^{r}\epsilon(r)~r^2 \rmd r.
\end{equation}
\noindent Here $\epsilon$ is the energy density and $m(r)$ denotes the mass enclosed by the radius $r$. The total gravitational mass of the configuration is given by $M=m(R)$. Given a value of the central energy density $\epsilon_{\mathrm{c}}$, the non-rotating configuration is determined by integrating the Tolman-Oppenheimer-Volkoff (TOV) equation of hydrostatic equilibrium,
\begin{equation}
\frac{\rmd p}{\rmd r}=-(\epsilon+p)\frac{m(r)+4\pi pr^3}{r[r-2m(r)]},
\end{equation}
together with the equation for the time metric component $e^{\nu(r)}$ given by
\begin{equation}
\frac{\rmd\nu}{\rmd r}=-\frac{2}{(\epsilon+p)}\left(\frac{\rmd p}{\rmd r}\right).
\end{equation}
\noindent In the exterior vacuum spacetime $\epsilon=p=0$, thus the geometry is described by the Schwarzschild metric
\begin{equation}\label{ext_schw}
 \rme^{\nu(r)}=\rme^{-\lambda(r)}=1-\frac{2M}{r},\quad r>R. 
\end{equation}
The interior and exterior geometries are matched across the boundary $\Sigma_{0}=R$
\begin{equation}
[\lambda]=0,\quad [\nu]=0,\quad [\nu']=0,
\end{equation}
\noindent where $[f]$ indicates the difference between the value of $f$ in the vacuum exterior and its value in the interior, evaluated at $\Sigma_0$, i.e., $[f]=f^{+}\vert_{\Sigma_0}-f^{-}\vert_{\Sigma_0}$, and the prime denotes a derivative with respect to the radial coordinate $r$ \footnote{We use the opposite convention of \cite{Reina:2015jia}, who uses $+ (-)$ to denote quantities in the interior (exterior).}.

\subsection{Hartle-Thorne equations for the rotational metric deformations}

 Once the equilibrium configuration is set into slow rotation, the interior distribution and the spacetime geometry around it change. The appropriate line element for this situation is \cite{Hartle:1967he,Hartle:1968si}
\begin{eqnarray}\label{axialmetric}
& \rmd s^2 = -\rme^{\nu}\left[1+2h_{0}(r)+2h_{2}(r)P_{2}(\cos\theta)\right]\rmd t^2\nonumber \\
&+\rme^{\lambda}\left\{1+\frac{\rme^{\lambda}}{r}\left[2m_{0}(r) + 2m_{2}(r)P_{2}(\cos\theta)\right]\right\}\rmd r^2 \nonumber\\
&+r^2\left[1+2(v_{2}-h_{2})P_{2}(\cos\theta)\right]\left[\rmd\theta^2 +\sin^2\theta (\rmd\phi - \omega \rmd t)^2\right],
\end{eqnarray}
\noindent where $P_{2}(\cos\theta)$ is the Legendre polynomial of order 2; $(h_{0},h_{2},m_{0},m_{2},v_{2})$ are quantities of order $\Omega^2$; and $\omega(r)$ corresponds to the angular velocity of the local inertial frame, relative to a distant observer. The function $\omega(r)$ is associated with the \emph{dragging} effect. Note that in the non-rotating case, the metric \eref{axialmetric} reduces to \eref{statmetric}.

We assume the fluid, inside the configuration, rotating uniformly with four-velocity $u^{\mu}$ given by %\cite{Hartle:1968}
\begin{eqnarray}
u^{t} &= (-g_{tt} - 2\Omega g_{t\phi} - g_{\phi\phi}\Omega^2)^{-1/2} \nonumber \\
&= \rme^{-\nu/2}\left[1+\frac{1}{2}r^2\sin^2\theta(\Omega-\omega)^2\rme^{-\nu}-h_{0}-h_{2}P_{2} \right], \nonumber\\
u^{\phi} &= \Omega u^{t},\quad u^{r}=u^{\theta}=0 \nonumber.
\end{eqnarray}
\noindent It is conventional to introduce the quantity 
\begin{equation}
\varpi\equiv \Omega - \omega,
\end{equation}
\noindent which corresponds to the angular velocity of the fluid relative to the local inertial frame. This quantity determines the centrifugal force and, at first order in $\Omega$, satisfies the following equation
\begin{equation}\label{omegain}
\frac{\rmd}{\rmd r}\left[r^4j(r)\frac{\rmd\varpi}{\rmd r}\right]+4r^3\frac{\rmd j}{\rmd r}\varpi=0,\quad j(r)\equiv \rme^{-(\lambda+\nu)/2}.
\end{equation}
In the region $r>R$, $\epsilon=p=0$ and the space-time geometry is described by \eref{ext_schw}. Thus, $j(r)=1$ and~\eref{omegain} can be easily integrated to give
\begin{equation}\label{omegaout}
\varpi(r)^{+}=\Omega - \frac{2J}{r^3},
\end{equation}
\noindent where $J$ is an integration constant associated with the angular momentum of the star \cite{Hartle:1967he}. For the interior solution, \eref{omegain} is integrated outward from the origin, with the boundary conditions $\varpi(0)=\varpi_{\mathrm{c}}=\mathrm{const}.$, and $(\rmd\varpi/\rmd r)\vert_{r=0}=0$. Regularity demands that, at the surface, the interior and exterior solutions, together with their derivatives, must match, i.e., $[\varpi]=[\varpi']=0$. Thus, one numerically integrates ~\eref{omegain} and obtains the surface value $\varpi(R)$, then one can determine the angular momentum $J$ and the angular velocity $\Omega$ as follows
\begin{equation}\label{momentumatR}
J=\frac{1}{6}R^4\left(\frac{\rmd\varpi}{\rmd r}\right)_{r=R},\quad \Omega = \varpi(R) + \frac{2J}{R^3}.
\end{equation}
Once the angular momentum and angular velocity are determined, the relativistic moment of inertia can be obtained from the relation $I = J/\Omega$.
\subsection{Spherical deformations: $l=0$ sector}
The spherical deformations are determined by the $l=0$ equations, for the perturbations $m_{0}$ and $p_{0}^{*}$, which read
\begin{eqnarray}\label{m0int}
\frac{\rmd m_{0}}{\rmd r}=4\pi r^2(\epsilon+p)\left(\frac{\rmd\epsilon}{\rmd p}\right)p_{0}^{*}\, + \frac{1}{12}r^4j^2\left(\frac{\rmd\varpi}{\rmd r}\right)^2-\frac{1}{3}r^3\varpi^2\frac{\rmd j^2}{\rmd r},
\end{eqnarray}
\begin{eqnarray}\label{h0int}
\frac{\rmd p_{0}^{*}}{\rmd r}&=&-\frac{1+8\pi p r^2}{(r-2m)^2}m_{0}-\frac{4\pi(\epsilon+p)r^2}{r-2m}p_{0}^{*}+\frac{1}{12}\frac{r^4 j^2}{(r-2m)}\left(\frac{\rmd\varpi}{\rmd r}\right)^2 \nonumber\\
&& +\frac{1}{3}\frac{\rmd}{\rmd r}\left(\frac{r^3 j^2 \varpi^2}{r-2m} \right).
\end{eqnarray}
These equations are integrated outward, from the origin, with the boundary conditions $h_{0}(0)=m_{0}(0)=0$. In the exterior of the star, \eref{m0int} and \eref{h0int} are integrated analytically giving
\begin{eqnarray}\label{m0out}
m_{0}^{+}(r)&=&\delta M - \frac{J^2}{r^3},\\
h_{0}^{+}(r)&=&-\frac{\delta M}{r-2M} + \frac{J^2}{r^3(r-2M)},
\end{eqnarray}
\noindent where $\delta M$ is an integration constant that is associated with the change in mass induced by the rotation which can be found by the matching of the solutions at the surface $\Sigma_{0}$, i.e., $[m_{0}]=[h_{0}]=0$. 

As pointed out by \cite{Reina:2014fga}, expression \eref{m0out} for $\delta M$ is not the most general one. By reconsidering the matching problem in the HT framework, by employing the perturbed matching theory at second order developed by \cite{Mars:2005ca}, \cite{Reina:2014fga} showed that the perturbative functions are continuous across the surface $\Sigma_{0}$, except for the function $m_{0}(r)$ in the $l=0$ sector. The discontinuity in $m_{0}(r)$ turns out to be proportional to the energy density at $\Sigma_{0}$, thus contributing an additional term to \eref{m0out} in the form
\begin{equation}\label{h0RV}
\delta M_{\mathrm{mod}} =m_{0}(R) + \frac{J^2}{R^3} + 8\pi R^3\left(\frac{R}{2M}-1\right)\epsilon(R)p_{0}^{*}.
\end{equation}
Note that for configurations with a vanishing surface energy density, as is the case for most of the EOS for realistic NSs including the T-VII solution, this additional term vanishes; thus the original HT expression \eref{m0out} remains valid. However, for configurations with a discontinuity in $\epsilon$ at the surface, e.g. constant density stars or strange stars \cite{Alcock:1986apj}, this additional term provides a significant contribution to the change in mass \cite{Reina:2015jia}.
\subsection{Quadrupole deformations: $l=2$ sector}
The quadrupole deformations of the star are determined by the $l=2$ perturbation equations given by
\begin{equation}\label{v2in}
\frac{\rmd v_{2}}{\rmd r}= -\left(\frac{\rmd\nu}{\rmd r}\right)h_{2} +\left(\frac{1}{r}+\frac{1}{2}\frac{\rmd\nu}{\rmd r}\right)\left[\frac{1}{6}r^4j^2\left(\frac{\rmd\varpi}{\rmd r}\right)^2-\frac{1}{3}r^3\varpi^2\frac{\rmd j^2}{\rmd r}\right].\\
\end{equation}
\begin{eqnarray}\label{h2in}
\frac{\rmd h_{2}}{dr}=\left\{-\frac{\rmd\nu}{dr}+\frac{r}{\left(r-2m\right)}\left(\frac{\rmd\nu}{\rmd r}\right)^{-1}\left[8\pi(\epsilon+p)-\frac{4m}{r^3}\right]\right\}h_{2}\nonumber\\
 -\frac{4v_{2}}{r\left(r-2m\right)}\left(\frac{\rmd\nu}{\rmd r}\right)^{-1}  + \frac{1}{6}\left[\frac{r}{2}\left(\frac{\rmd\nu}{\rmd r}\right)-\frac{1}{\left(r-2m\right)}\left(\frac{\rmd\nu}{\rmd r}\right)^{-1}\right]r^3j^2\left(\frac{\rmd\varpi}{\rmd r}\right)^2 \nonumber\\
-\frac{1}{3}\left[\frac{r}{2}\left(\frac{\rmd\nu}{\rmd r}\right)+\frac{1}{\left(r-2m\right)}\left(\frac{\rmd\nu}{\rmd r}\right)^{-1}\right]r^2\frac{\rmd j^2}{\rmd r}\varpi^2,
\end{eqnarray}
\noindent 
The solutions of \eref{v2in} and \eref{h2in} can be expressed as the sum of a homogeneous part and a particular solution
\begin{equation}
h_{2} = Ah_{2}^{\mathrm{h}} + h_{2}^{\mathrm{p}},\quad v_{2} = Av_{2}^{\mathrm{h}} + v_{2}^{\mathrm{p}},
\end{equation}
\noindent where the particular solution is given by the integration of \eref{v2in} and \eref{h2in}, with the following near-the-origin behaviours
\begin{equation} 
h_{2}^{\mathrm{p}}=ar^2,\quad v_{2}^{\mathrm{p}}=br^2,
\end{equation}
where, for an arbitrary value of $a$, the constant $b$ satisfies the following constraint \cite{Chandra:1974, Miller:1977}
\begin{equation} 
b=\frac{2\pi}{3}\left[(\rhoc+p_{\mathrm{c}})j_{\mathrm{c}}^2-(\rhoc+3p_{\mathrm{c}})a\right].
\end{equation}
On the other hand, the homogenous integrals are obtained from the following equations
\begin{equation}\label{v2_hom}
\frac{\rmd v_{2}^{\rm{h}}}{\rmd r}= -\left(\frac{\rmd\nu}{\rmd r}\right)h_{2}^{\mathrm{h}},
\end{equation}
\begin{eqnarray}\label{h2_hom}
\frac{\rmd h_{2}^{\rm{h}}}{\rmd r} &=& - \left\{\frac{\rmd\nu}{\rmd r} -\frac{r}{\left(r-2m\right)}\left(\frac{\rmd\nu}{\rmd r}\right)^{-1}\left[8\pi(\epsilon+p)-\frac{4m}{r^3}\right]\right\}h_{2}^{\rm{h}}\nonumber\\
&&-\frac{4v_{2}^{\rm{h}}}{r\left(r-2m\right)}\left(\frac{\rmd\nu}{\rmd r}\right)^{-1}.
\end{eqnarray}
These equations are integrated outward from the centre of the star where they behave as follows
\begin{equation}
h_{2}^{\mathrm{h}}=Br^2,\quad v_{2}^{\mathrm{h}}=-\frac{2\pi}{3}\left(\rhoc+3p_{\mathrm{c}}\right)Br^4,
\end{equation}
\noindent where $B$ is an arbitrary constant. In the vacuum exterior, \eref{v2in} and \eref{h2in} can be integrated analytically to give
\begin{equation}\label{h2_ext}
h_{2}^{+}(r)=J^2\left(\frac{1}{Mr^3}+\frac{1}{r^4}\right)+KQ_{2}^{\;2}\left(\frac{r}{M}-1\right),
\end{equation}
\begin{equation}\label{v2_ext}
v_{2}^{+}(r)=-\frac{J^2}{r^4}+K\frac{2M}{\left[r(r-2M)\right]^{1/2}}Q_{2}^{\;1}\left(\frac{r}{M}-1\right),
\end{equation}
\noindent where $K$ is an integration constant, and $Q_{n}^{\;m}$ are the associated Legendre functions of the second kind. The condition of regularity of the solutions at the surface $\Sigma_{0}$ demands $[h_2]=[v_2]=0$. The constant $K$ appearing in Eqs.~\eref{h2_ext} and \eref{v2_ext}, is related to the mass quadrupole moment of the star, as measured at infinity, via the relation
\begin{equation}\label{quadrupole}
Q = \frac{J^2}{M} + \frac{8}{5}KM^3.
\end{equation}
To find the constant $K$, the functions $h_2$ and $v_2$ must be computed for the interior of the star, and then matched at the surface $\Sigma_0$ with the exterior solutions \eref{h2_ext} and \eref{v2_ext}.

The perturbations for mass and pressure $m_{2}$ and $p_{2}^{*}$, which determine the rotational deformations of the star, can be determined using the following relations \cite{Hartle:1968si}
\begin{eqnarray}\label{m2_in}
m_{2}=\left(r-2m\right)\left[-h_{2}-\frac{1}{3}r^3\left(\frac{\rmd j^2}{\rmd r}\right)\varpi^2 +\frac{1}{6}r^4j^2\left(\frac{\rmd\varpi}{\rmd r}\right)^2\right],
\end{eqnarray}
\begin{equation}\label{p2_in}
p_{2}^{*} = -h_{2} - \frac{1}{3}r^2 \rme^{-\nu} \varpi^2.
\end{equation}
The equation that describes the isobaric surfaces can be written as \cite{Chandra:1974, Miller:1977}
\begin{equation}\label{radio}
r(p)=r_{0} + \xi_{0}(r) + \xi_{2}(r)P_{2}(\cos\theta),
\end{equation}
\noindent where $r_{0}$ is the radius of the nonrotating configuration, and where the pressure has value $p$. The deformations are given by
\begin{equation}\label{defxi0}
\xi_{0}(r)=-p_{0}^{*}(\epsilon+p)\left(\frac{\rmd p}{\rmd r}\right)^{-1},
\end{equation}
\begin{equation}\label{defxi2}
\xi_{2}(r)=-p_{2}^{*}(\epsilon+p)\left(\frac{\rmd p}{\rmd r}\right)^{-1}.
\end{equation}
Equation~\eref{radio} is written in a particular coordinate system. However, one can construct an invariant parametrization of the isobaric surface by embedding it in a three-dimensional flat space with polar coordinates $r^{*}, \theta^{*}, \phi^{*}$ which has the same intrinsic geometry as the isobaric surface. Thus, the corresponding $3$-surface in flat space (at order $\Omega^2$) is given by
\begin{equation}
r^{*}(\theta^{*})=r_{0} + \xi_0(r) + [\xi_2 + r(v_2 - h_2)]P_{2}(\cos\theta).
\end{equation}
\noindent The ellipticity of this spheroid is given by \cite{Miller:1977}
\begin{equation}\label{ellipt}
\varepsilon=-\frac{3}{2r}\left[\xi_2 + r(v_2 - h_2)\right].
\end{equation}
Miller \cite{Miller:1977} has shown that there is an alternative way to define ellipticity using proper distances. However, in the following, we use the definition given by \eref{ellipt}. 

%%%%%%%%%%%%%%%%%%%%%%%%%%%%%%%

\section{Tolman VII fluid spheres}\label{sec:Tolman}

In this section, we review the T-VII solution \cite{Tolman:1939,Lattimer:2000nx,Posada:2021zxk}. In his prescription, Tolman assumed the radial metric component $g_{rr}=\rme^{\lambda(r)}$ as being in the form
\begin{equation}\label{grrTol}
  \rme^{-\lambda(r)} = 1 - \frac{\xi^2}{2\beta} (5 - 3\xi^2)\,,
\end{equation}
where $\xi\equiv r/R$ is the radial coordinate normalized by the stellar radius, and $\beta\equiv R/\Rs$, where $\Rs(= 2M)$ is the Schwarzschild radius, denotes the ‘tenuity' parameter \cite{Neary:2001ai}. In contrast with \cite{Lattimer:2000nx,Posada:2021zxk}, instead of using the compactness $\mathcal{C}\equiv M/R$, here we parametrize the T-VII solution by $R/\Rs$. These quantities are connected via the relation $\com=1/(2\beta)$. 

The remaining functions describing the T-VII solution are given as follows: the energy density $\epsilon$ follows a quadratic fall-off form, as a function of $r$, given by
\begin{equation}\label{rhoTol}
  \epsilon(\xi) = \rhoc (1 - \xi^2),
\end{equation}
\noindent where $\rhoc$ indicates the central energy density. Note that $\epsilon$ vanishes at the surface $\xi=1$. From \eref{misner-sharp} we obtain the mass enclosed by the radius $r$ as
\begin{equation}\label{mTol}
m(\xi) = \frac{M}{2} \xi^3 (5 - 3\xi^2),
\end{equation}
\noindent which gives the total gravitational mass $M$ when evaluated at the surface $\xi=1$. On the other hand, the $g_{tt}$ metric component is given by
\begin{equation}\label{gttTol}
  \rme^{\nu} = C_1 \cos^2 \phi(\xi)\,,
\end{equation}
where
\begin{equation}\label{phiTol}
  \phi(\xi) = C_2 - \frac{1}{2}\log \left(\xi^2 - \frac{5}{6} + \sqrt{\frac{2\beta}{3\rme^{\lambda}}}\right),
\end{equation}
\noindent and $C_1$ and $C_2$ are integration constants given by
\begin{eqnarray}%\label{C1Tol}
C_{1} &= 1 - \frac{5}{6\beta}\nonumber,\\
C_{2} &= \arctan{\sqrt\frac{1}{6(\beta-1)}} + \frac{1}{2}\log\left(\frac{1}{6}+\sqrt{\frac{2(\beta-1)}{3}}\right)\nonumber.
\end{eqnarray}
Finally, the radial pressure is found to be
 \begin{equation}\label{pTol}
  p(\xi) = \frac{\rhoc}{15} \left[\sqrt{24\beta \rme^{-\lambda}}\tan\phi - (5 - 3\xi^2)\right]\,,
\end{equation}
\noindent which vanishes at the surface $\xi=1$. With the help of \eref{rhoTol} and \eref{mTol}, we can find the relation between $\beta$ and the central energy density as being
\begin{equation}
\rhoc=\frac{15}{16\pi\beta R^2}.
\end{equation}

\begin{figure}%[h]
\centering
\includegraphics[width=0.6\linewidth]{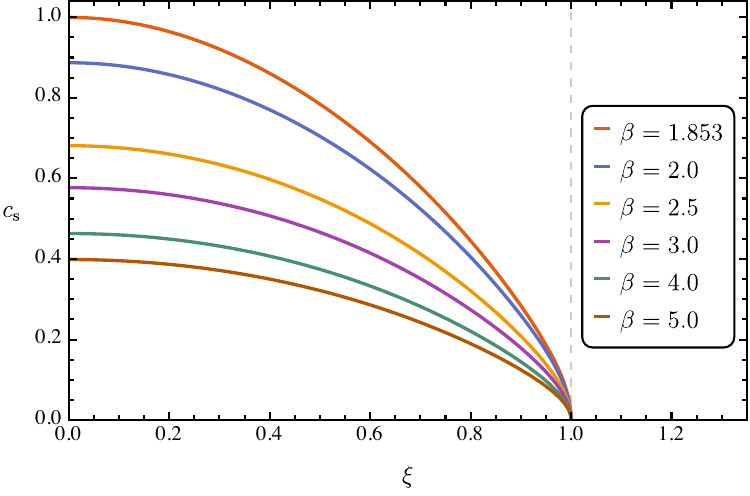}
\caption{Profiles of the speed of sound $\cs=\sqrt{\left(\partial p/\partial\epsilon\right)}$, as a function of the radial coordinate $\xi$, for the T-VII solution. The various curves shown here correspond to different values of the tenuity parameter $\beta$.}
\label{fig:cs}
\end{figure}

Let us recall certain restrictions, on the parameter $\beta$, for the physical plausibility of the T-VII solution. The central pressure is finite for $\beta > 1.2946$, which is above the lower value allowed by the Buchdahl theorem (valid for any EOS), namely, $\beta\geq 9/8$ \cite{Buchdahl:1959zz}. On the other hand, the dominant energy condition (DEC), which demands that the pressure should not exceed the energy density, i.e., $\epsilon>\vert p\vert$, is valid for $\beta>1.4921$ \cite{Posada:2021zxk}. Causality, as determined by the condition that the central speed of sound, $\cs(0)=\sqrt{\left(\partial p/\partial\epsilon\right)}\vert_{\xi=0}\leq1$, is valid up to $\beta=1.8540$. In \fref{fig:cs} we show the profiles of the speed of sound $\cs$, for different values of the parameter $\beta$, as a function of the radial coordinate $\xi$. Note that $\cs$ is maximum at the centre and decreases monotonically with $r$, going to zero at the surface. Additionally, the T-VII solution is stable, against radial oscillations, for $\beta>1.4585$ \cite{Moustakidis:2016ndw,Posada:2021zxk}. We summarize these bounds in \tref{tab:1}. 

\begin{table}%[h]
\caption{\label{tab:1}Limits on the tenuity $\beta$ for the physical viability of the T-VII spheres, as determined by the following conditions: (a) finite ratio $p_{\mathrm{c}}/\rhoc$, (b) radial stability, (c) dominant energy condition, and (d) causality.}
\begin{indented}
\item[]\begin{tabular}{@{}ll}
 \br
 Condition & $\beta$ \\
 \mr
 Finite $p_{\mathrm{c}}/\rhoc$ & $1.2946$ \\
 Stability & $1.4585$ \\
 DEC & $1.4921$ \\
 $\cs(0)\leq 1$ & $1.8532$ \\
 \br
\end{tabular}
\end{indented}
\end{table}
%%%%%%%%%%%%%%%%%%%%%%%%%%%%%%
\section{Numerical results for slowly rotating T-VII spheres}\label{sec:Results}

In this section, we present the results of our numerical integrations of the HT structure equations, for slowly rotating T-VII fluid spheres, up to the second order in the angular velocity $\Omega$. We consider a sequence of models representing an adiabatic and quasi-stationary contraction by progressively reducing the radius while keeping the angular momentum $J$ and gravitational mass $M$ constant; we integrated the structure equations for a sequence of values of $R/\Rs$, up to the limiting value for dynamical stability $\beta_{\mathrm{stab}}$ (see~\tref{tab:1}), using a standard fourth-order Runge-Kutta algorithm. Although the limits on $\beta$ as given by the DEC and sub-luminal central speed of sound are stricter than those given by the dynamical stability condition, we consider the latter as our limiting $\beta$ during the integrations. In \tref{tab:2} we list some of the properties of the models evaluated at the surface. The main results of the integrations are presented further in figures \ref{fig:drag_rad}-\ref{fig:q}. Following \cite{Chandra:1974, Miller:1977} we use dimensionless variables where the units in which the quantities are expressed, are given in the corresponding descriptions.

\subsection{Dragging of inertial frames and moment of inertia}

\begin{figure}
\centering
\includegraphics[width=0.495\linewidth]{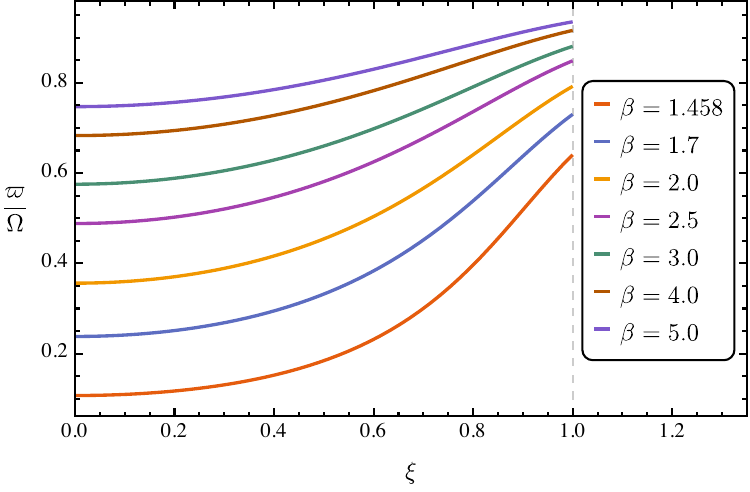}
\includegraphics[width=0.495\linewidth]{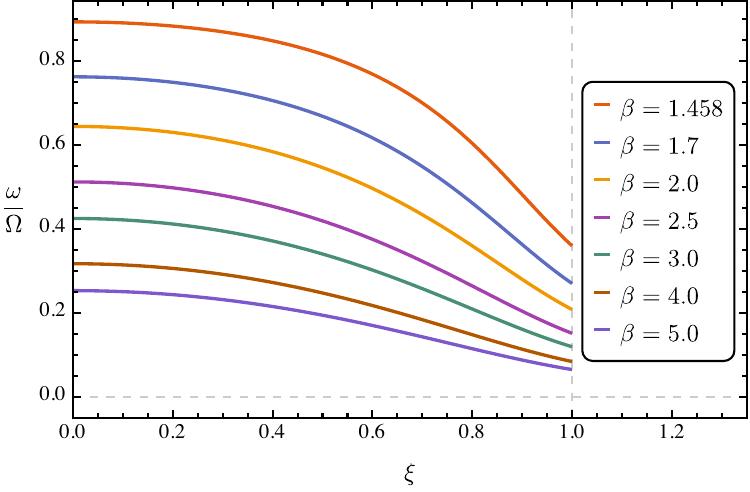}
\caption{\emph{Dragging of the inertial frames}. {\bf Left panel:} radial profile of the angular velocity $\varpi/\Omega=(1-\omega/\Omega)$, relative to the local inertial frame (as measured by a distant observer), for the T-VII solution. We normalize $\varpi$ by the fluid's angular velocity $\Omega$ relative to a distant observer. The curves are labelled by their corresponding values of $\beta$. {\bf Right panel:} angular velocity $\omega/\Omega$, or dragging of the inertial frames, as a function of the radial coordinate $\xi$. We show profiles for the same values of $\beta$ as in the right-hand panel. We normalize $\omega$ by the fluid's angular velocity $\Omega$ relative to a distant observer. Note that the dragging is maximum at the centre of the star, and it decreases outwards.}
\label{fig:drag_rad}
\end{figure}

The dragging of the inertial frames, in the T-VII space-time, is shown~\fref{fig:drag_rad}. In the left-hand panel, we plot the radial profile of $\varpi(r)$, corresponding to the angular velocity of the fluid relative to the local inertial frame, for different values of the parameter $\beta$. We observe that $\varpi$ has a local minimum at the origin, and it grows monotonically with $r$ up to the surface $\xi=1$. Therefore, the dragging of inertial frames, as measured by a distant observer, is greatest at the centre of the star and decreases outwards, as we also show in the right-hand panel of the same figure. We also observe that as $\beta$ decreases, i.e. as the star becomes more compact, the $\varpi(r)$ function is reduced, while the ‘dragging function' $\omega(r)$ is enhanced; the more compact the configuration, the greater the dragging of local inertial frames in the interior of it. For the limiting value $\beta_\mathrm{stab}$, for dynamical stability, the central value of $\varpi$ is close to zero, but it does not vanish, in agreement with a theorem due to \cite{Hartle:1967he}.

\begin{figure}%[h]
\centering
\includegraphics[width=0.6\linewidth]{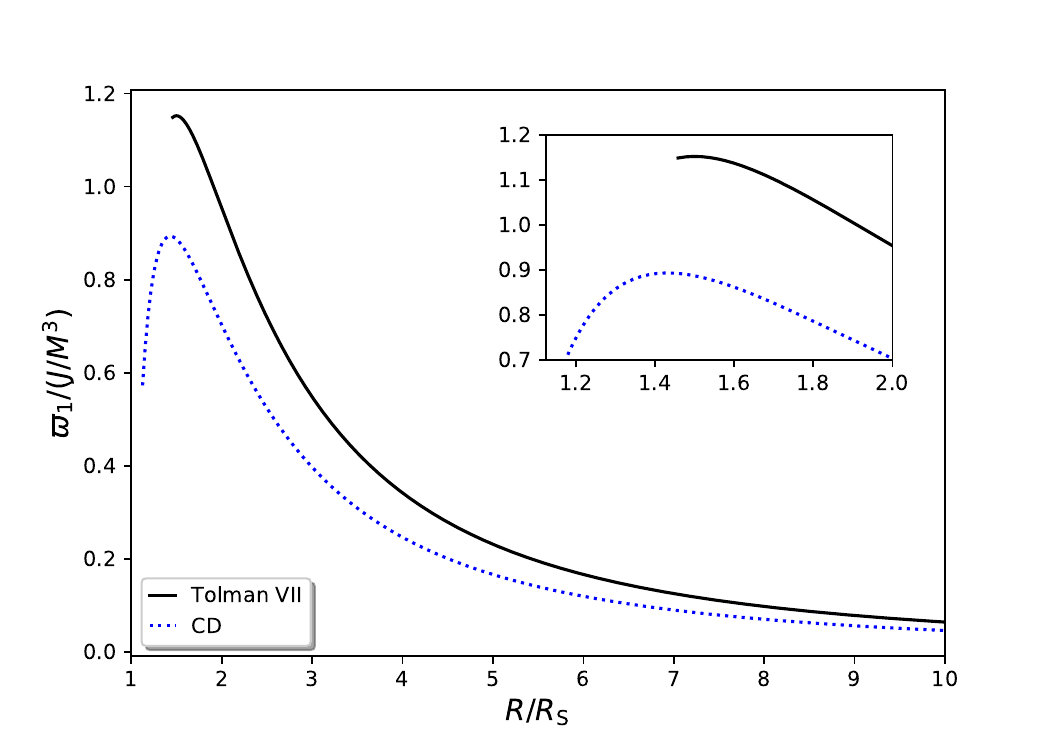}
\caption{Surface value of the angular velocity $\varpi_1$, relative to the local inertial frame, for different values of the parameter $R/\Rs$, for the T-VII spheres (solid black curve). We include the results for uniform-density stars (dotted blue curve). We measure $\varpi_1$ in units of $J/M^3$. The inset shows a magnification in the region near the local maxima.}
\label{fig:drag_surf}
\end{figure}

In \fref{fig:drag_surf} we present the surface value $\varpi_{1}=\varpi(R)$, as a function of $R/\Rs$, for the T-VII solution (solid black curve). As a code test, we determined $\varpi_{1}$ for constant density (CD) configurations (blue dotted line), which are described by the Schwarzschild interior solution \cite{Schaffner-Bielich:2020psc}; we confirmed the results reported by Chandrasekhar and Miller \cite{Chandra:1974} with very good agreement. For the T-VII spheres, we observe that as the compactness of the star increases, $\varpi_1$ increases, reaching a local maximum at $\beta\simeq 1.5$ and then it decreases. We observe a similar behaviour for CD configurations, although the maximum there appears at $\beta \simeq 1.43$. Note that the Schwarzschild interior solution allows higher compactness, up to the Buchdahl limit $\beta=9/8$ where the central pressure diverges, in comparison with the T-VII solution.

\begin{figure*}%[h]
\centering
\includegraphics[width=0.495\linewidth]{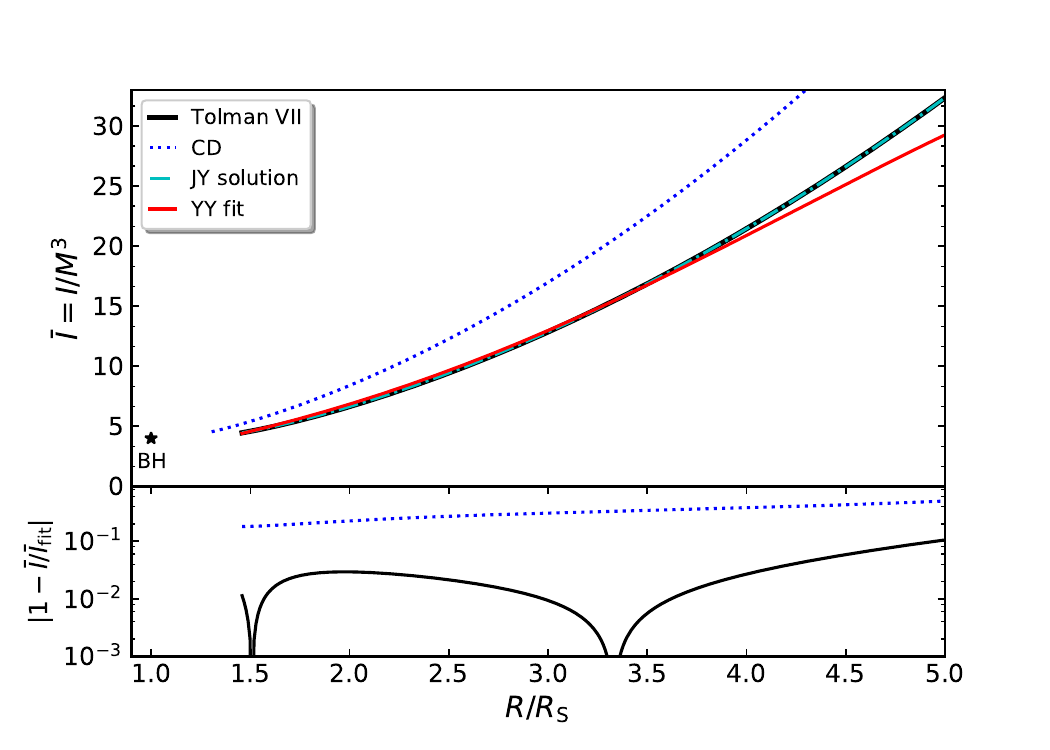}
\includegraphics[width=0.495\linewidth]{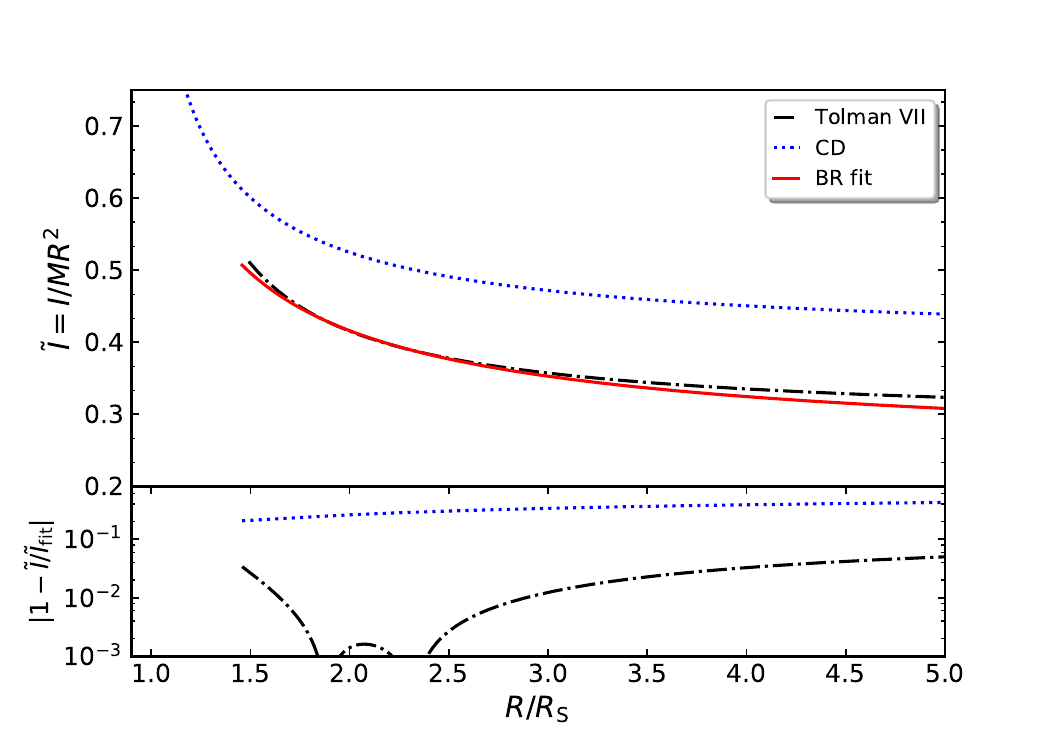}
\caption{\emph{Moment of inertia for the T-VII solution}. {\bf Left panel:} Dimensionless moment of inertia $\bar{I}=I/M^3$, as a function of $R/\Rs$, for the T-VII solution (solid black curve). The star indicates the black hole (BH) value $\bar{I}=4$. We include the numerical results for constant density (CD) configurations (dotted blue curve). We also plot the analytical solution for $\bar{I}$ found by Jiang and Yagi (JY) \cite{Jiang:2020uvb} for the T-VII solution (dotted cyan curve), as well as the Yagi-Yunes (YY) fit for the $\bar{I}-\mathcal{C}$ relation for NSs \eref{YYfit} (solid red curve). The bottom panel shows the fractional differences between the numerical results and the YY fit. {\bf Right panel}: normalized moment of inertia $\tilde{I}=I/(MR^2)$ plotted as a function of the parameter $R/\Rs$, for the T-VII solution (dashed black curve). We include the numerical results obtained for CD configurations (dotted blue line). We also plot the polynomial fit proposed by Breu and Rezzolla (BR) \cite{Breu:2016ufb} for realistic EOS of NSs (solid red curve). The bottom panel shows the relative error between the numerical results and the BR fit.}
\label{fig:inertia}
\end{figure*}

The left-hand panel of~\fref{fig:inertia} displays the dimensionless moment of inertia $\bar{I}\equiv I/M^3$, plotted as a function of $\beta$. The quantity $\bar{I}$ has been shown to be relevant in the context of the $I$-Love-$Q$ \cite{Yagi:2014qua,Yagi:2013awa}, and $I$-Love-$\com$ relations \cite{Jiang:2020uvb}. We observe that $\bar{I}$ decreases monotonically, and seems to approach the black hole (BH) limit $\bar{I}_{\mathrm{\tiny BH}}=4$, as $\beta\to 1$. For instance, for the limiting value for stability $\beta_{\mathrm{stab}}$ we obtain $\bar{I}=4.460$. Notice however that the strict BH limit ($\beta\to 1$), cannot be taken from this solution, or any other EOS for realistic NSs, for any finite value of the central energy density \footnote{It has been shown that, under certain conditions, uniform density stars can evade the Buchdahl bound and approach arbitrarily close to the BH compactness limit where they become essentially a gravastar \cite{Mazur:2015kia, Posada:2018agb}. Moreover, in this limit, quantities like the Love number, normalized moment of inertia, and quadrupole moment, approach the corresponding Kerr values \cite{Chirenti:2020bas, Beltracchi:2021lez}.}. In the same figure, we also include results of $\bar{I}$ for homogeneous configurations.  We observe that the CD solution predicts a $\bar{I}-\mathcal{C}$ relation quite different from the one predicted by the T-VII model; however, as the compactness approaches the maximum value for stability, the corresponding values of $\bar{I}$ seem to approach to the BH value $\bar{I}_{\mathrm{\tiny BH}}=4$.

In order to compare with the $\bar{I}-\mathcal{C}$ relations for realistic NSs, in the same figure we include the polynomial fit proposed by Yagi and Yunes \cite{Yagi:2016bkt} in the form

\begin{equation}\label{YYfit}
\bar{I}=\sum_{k=1}^{4}a_{k}\mathcal{C}^{-k},
\end{equation}

\noindent where the best-fitted coefficients are: $a_{1}=1.317$, $a_{2}=-0.05043$, $a_{3}=0.04806$ and $a_{4}=-0.002692$. We also plot the analytical $\bar{I}-\mathcal{C}$ relation found by Jiang and Yagi \cite{Jiang:2020uvb} for the T-VII solution. The bottom panel shows the relative fractional difference between our numerical results for the CD and T-VII solution, and the fit~\eref{YYfit}. First of all, we observe the excellent agreement between our results and the analytical Jiang-Yagi solution \cite{Jiang:2020uvb} for the T-VII solution. We also note that the T-VII solution agrees well with the fit~\eref{YYfit} in the range of $\beta\sim 1.45-3.8$ where the relative differences are below $2\%$, while the maximum fractional difference is at most $10\%$, which is consistent with the results found for realistic EOSs \cite{Breu:2016ufb}. On the other hand, we observe that the $\bar{I}-\mathcal{C}$ relation for CD configurations differs from the one predicted by the T-VII solution, as well as the one for NSs in agreement with previous results \cite{Yagi:2016bkt}.

It has been suggested by some authors \cite{Lattimer:2000nx,Ravenhall:1994apj,Bejger:2002ty,Urbanec:2013fs} that the normalized moment of inertia $\tilde{I}\equiv I/MR^2$, for stars in slow rotation, can be expressed as a function of the compactness of the configuration $\com\equiv M/R$, through some relatively simple low-order polynomial functions which have a mild dependence on the EOS. For instance, by considering a number of EOSs for realistic NSs, \cite{Lattimer:2004nj} proposed a quartic order polynomial fit given by 
\begin{equation}\label{fit}
\tilde{I}=\tilde{a}_{0}+\tilde{a}_{1}\com+\tilde{a}_{4}\com^4,
\end{equation}
\noindent with the following fitting coefficients, $\tilde{a}_{0}=0.237\pm 0.008$, $\tilde{a}_{1}=0.674$ and $\tilde{a}_{4}=4.48$. More recently, \cite{Breu:2016ufb} considered a larger number of EOS for realistic NSs in a wide range of compactness, namely, $\com\in[0.07,0.32]$. They found that the fitting coefficients are $\tilde{a}_{0}=0.244$, $\tilde{a}_{1}=0.638$ and $\tilde{a}_{4}=3.202$. 

In the right-hand panel of~\fref{fig:inertia} we plot the same numerical results for the T-VII solution as in the left-hand panel but now normalized as $I/MR^2$, for various values of $R/\Rs$. Note that as $\beta$ decreases, or the compactness increases, $\tilde{I}$ grows monotonically approaching a value above $0.5$, when $\beta$ approaches the limiting value for dynamical stability. We also include our numerical results for CD configurations, which are in very good agreement with those reported by \cite{Chandra:1974}. Here we observe that the values of $I/MR^2$ predicted by CD configurations are quite different from those predicted by the T-VII solution, as well as those for NSs. 

In order to compare the predictions for $\tilde{I}$ from the T-VII solution with those using realistic EOSs for NSs, we include the fitting polynomial formula \eref{fit} with the coefficients reported by \cite{Breu:2016ufb}. In the bottom panel, we show the relative error. We observe that the greatest differences appear for $\beta>3.5$ where the relative errors are above $5\%$, while the agreement improves significantly in the range $1.63<\beta<2.9$, which roughly speaking is the relevant range of compactness for realistic NSs \cite{Miller:2021qha}, where the relative error is below $1\%$. Thus, these results suggest that, in terms of the normalized moment of inertia, the T-VII solution provides a very good description of a realistic NS in the relevant regime.

\subsection{Spherical deformations: $l=0$ sector}

\begin{figure}%[h!]
\centering
\includegraphics[width=0.6\linewidth]{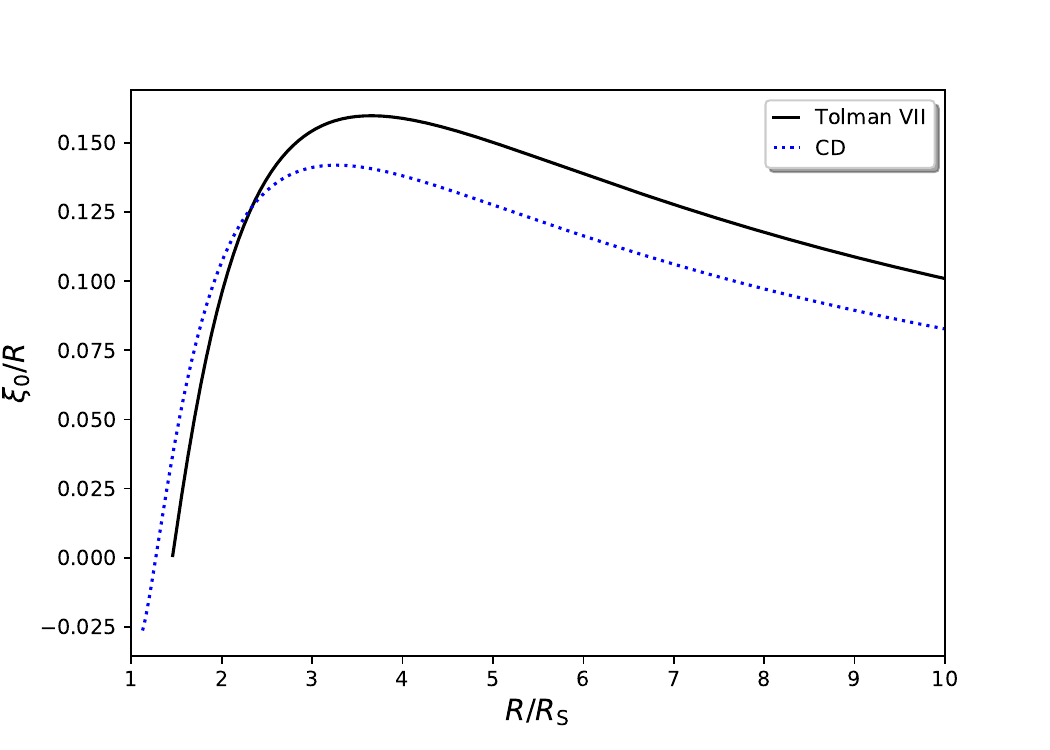}
\caption{The $l=0$ deformation of the surface $\xi_0/R$ (measured in units of $J^2/M^4$) as a function of $R/\Rs$, for T-VII spheres (solid black curve) and CD configurations (dotted blue curve).}
\label{fig:xi0}
\end{figure}

In~\fref{fig:xi0} we plot the surface value of the deformation parameter $\xi_{0}/R$~\eref{defxi0}, as a function of $R/\Rs$, for the T-VII solution (solid black curve). We also include results for CD configurations (dotted blue curve), which are in very good agreement with those reported by \cite{Chandra:1974}. For the T-VII spheres, we observe that $\xi_{0}$ is not monotonic; instead, as $\beta$ decreases $\xi_{0}$ increases reaching a local maximum at $\beta\simeq 3.65$, and then it decreases approaching zero as $\beta\to\stab$. For CD configurations we observe a similar behaviour, although the corresponding maximum of $\xi_{0}/R$ appears at $\beta\sim 3.27$. 

\begin{figure}%[h!]
\centering
\includegraphics[width=0.6\linewidth]{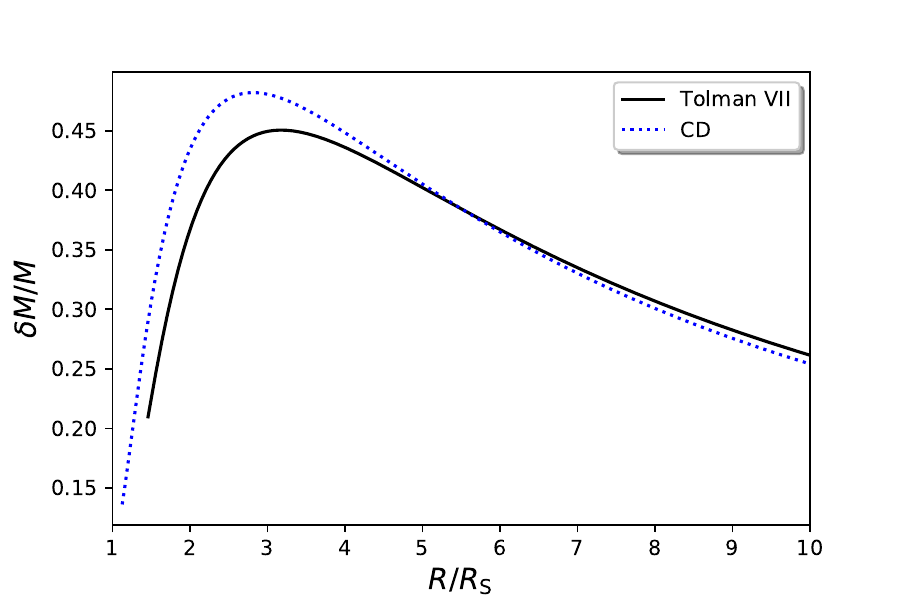}
\caption{The fractional change of mass $\delta M/M$ plotted against the parameter $R/\Rs$, for T-VII fluid spheres (solid black curve) and homogeneous masses (dotted blue curve); $\delta M/M$ is measured in units of $J^2/M^4$. }
\label{fig:deltaM}
\end{figure}

The fractional change of mass $\delta M/M$, as given by~\eref{h0RV}, is shown in~\fref{fig:deltaM} as a function of $R/\Rs$ for the T-VII solution (solid black curve) and the uniform density configurations (dotted blue curve). First of all, we confirmed the results for homogeneous masses reported by \cite{Reina:2015jia}. It is worthwhile to recall that in the T-VII solution, the energy density vanishes at the surface, thus the last term on the right-hand side of~\eref{h0RV} does not contribute to $\delta M$. We observe that $\delta M/M$ is not monotonic, instead, it grows as $\beta$ decreases reaching a maximum at $\beta\sim 3.18$, and then it decreases as $\beta\to\stab$. For constant density stars we observe that the function $\delta M/M$ shows a similar behaviour, although the maximum there appears at $\beta\sim 2.81$.  

\subsection{Quadrupole deformations: $l=2$ sector}

\begin{figure}%[h!]
\centering
\includegraphics[width=0.6\linewidth]{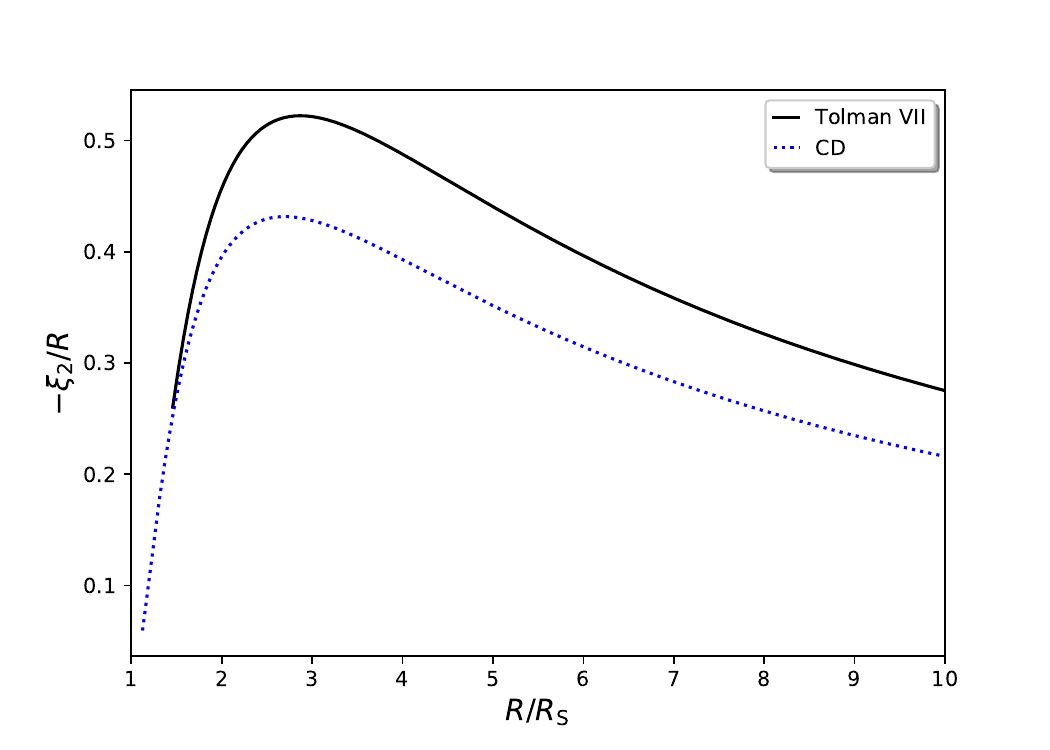}
\caption{The $l=2$ deformation of the surface $\xi_2(R)/R$ (measured in units of $J^2/M^4$) plotted as a function of $R/\Rs$, for the T-VII solution (solid black curve) and CD configurations (dotted blue curve).}
\label{fig:xi2}
\end{figure}

In~\fref{fig:xi2} we display the deformation function $-\xi_2(R)/R$, as given by~\eref{defxi2}, evaluated at the surface, as a function of the parameter $\beta$ for the T-VII solution (solid black curve). We also include the results for CD configurations (dotted blue curve), which are in very good agreement with the results reported by \cite{Chandra:1974}. In analogy to the behaviour of the $l=0$ perturbation function $\xi_{0}$, as $\beta$ decreases, or the compactness increases, $-\xi_2(R)/R$ increases reaching a maximum at $\beta\simeq 2.86$, and then decreases as $\beta\to\stab$. We observe a similar behaviour for homogeneous masses, with $-\xi_2(R)/R$ having a maximum when $\beta\sim 2.7$. 

\begin{figure}%[h!]
\centering
\includegraphics[width=0.6\linewidth]{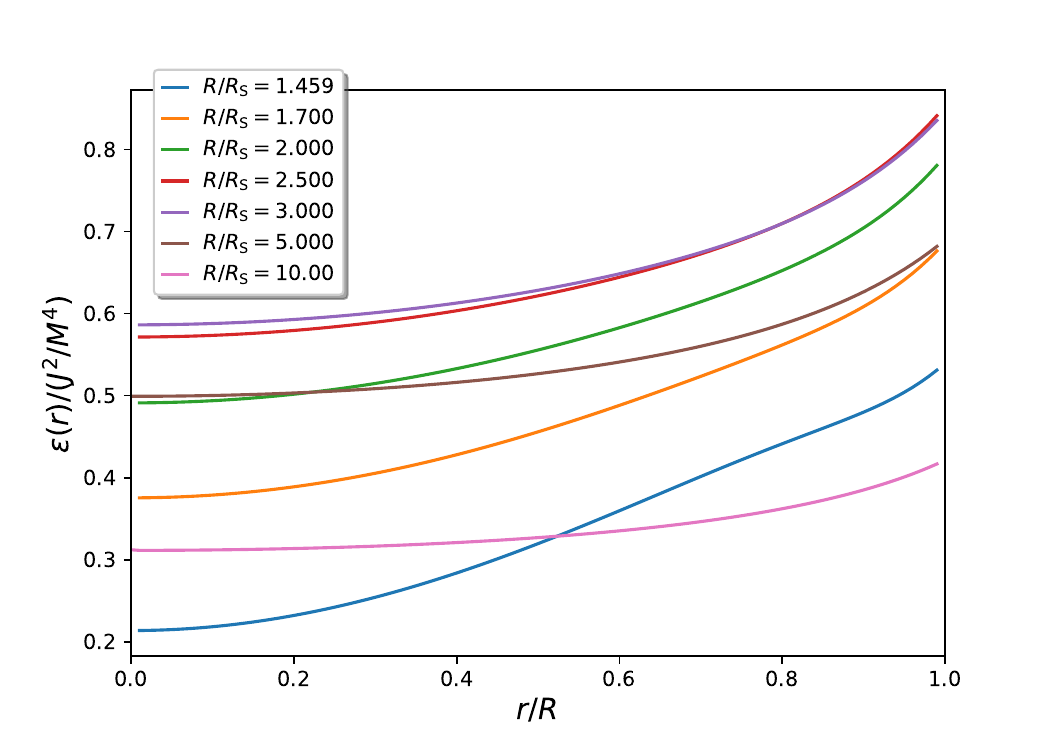}
\caption{Radial profile of the ellipticity $\varepsilon(r)$ for the T-VII solution, in units of $J^2/M^4$, for various values of $R/\Rs$.}
\label{fig:ellip_rad}
\end{figure}

The radial profile of the ellipticity $\varepsilon(r)$ of the isobaric surfaces, for T-VII spheres, as given by~\eref{ellipt}, is illustrated in~\fref{fig:ellip_rad} for various values of the parameter $R/\Rs$. We measure $\varepsilon(r)$ in units of $J^2/M^4$. We observe that the ellipticity grows monotonically from the centre of the star, up to the surface. Also, note that for configurations with low compactness, the region $r/R\in (0, 0.6)$ is less deformed as compared with the region $r/R\in (0.6, 1)$. For instance, for the configuration with $\beta=10$, equivalent to $\com=0.05$, the ellipticity is almost constant up to $r/R\sim 0.8$. On the other hand, for the configuration with the limiting $\beta$ for stability, we observe that the ellipticity grows almost linearly with $r$.

\begin{figure}%[h!]
\centering
\includegraphics[width=0.6\linewidth]{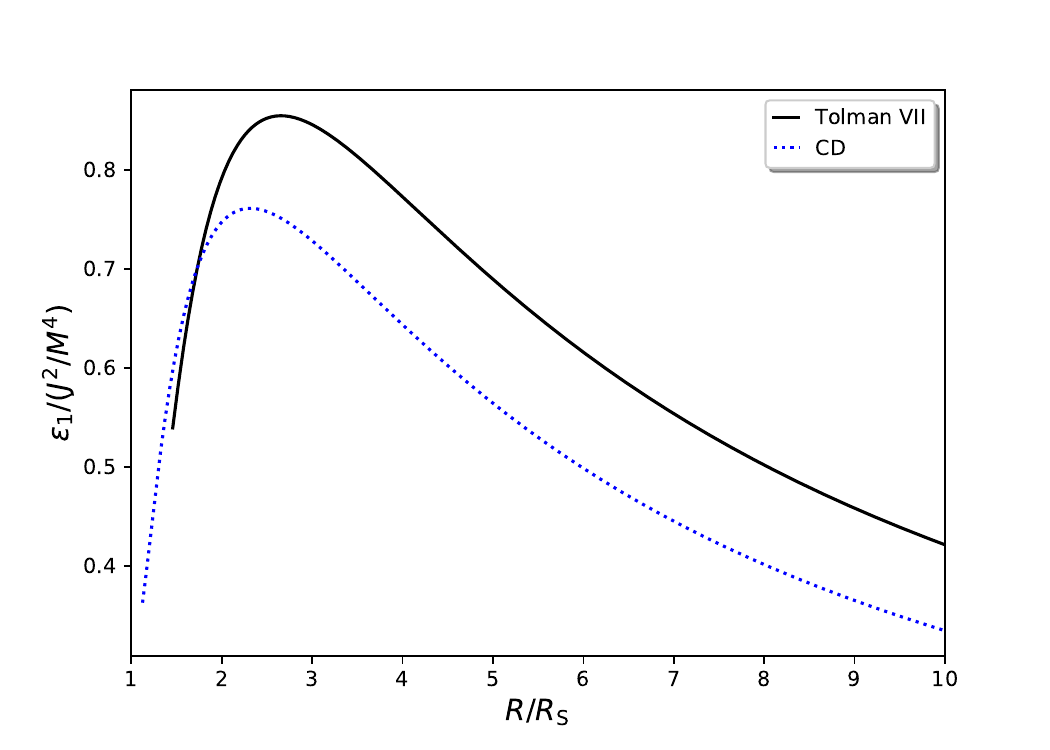}
\caption{The ellipticity of the surface $\varepsilon_{1}$ (in units of $J^2/M^4$), as a function of the parameter $R/\Rs$ for the T-VII solution (solid black curve) and homogeneous masses (dotted blue curve).}
\label{fig:ellip_surf}
\end{figure}

The ellipticity of the bounding surface $\varepsilon_{1}=\varepsilon(R)$, as a function of $R/\Rs$, is shown in~\fref{fig:ellip_surf} for the T-VII fluid spheres (black solid line). We also include the results for CD configurations (dotted blue line) which are in excellent agreement with the results reported by \cite{Chandra:1974}. For the T-VII model, we observe that the ellipticity first grows as the star increases its compactness while keeping mass and angular momentum fixed but, when $\beta\sim 2.65$ it reaches a maximum and then decreases. Thus, when a T-VII spheroid is contracted beyond this point, it will become more spherical rather than more oblate. We observe a similar behaviour in the ellipticity of uniform density configurations, where the maximum of $\varepsilon_{1}$ appears when $\beta\sim 2.31$, which corrects the value reported by \cite{Chandra:1974}. This reversal in the behaviour of the ellipticity, which is not predicted in Newtonian mechanics, was also found for polytropes \cite{Miller:1977} in slow rotation. Note that the turning $\beta$-point in the ellipticity, for the T-VII spheroids, is a little above the one corresponding to the maximum found for homogeneous stars, but it seems to be closer to the maximum found for polytropes with $\gamma=5/2$ \cite{Miller:1977}. In principle, our results support the hypothesis drawn by \cite{Miller:1977}, namely that for a contraction where $J/M^2$ remains fixed, a maximum in the ellipticity will occur for $\beta$ in the range 2 to 3 if the EOS allows high-density configurations to exist in hydrostatic equilibrium.

\begin{figure}%[h!]
\centering
\includegraphics[width=0.6\linewidth]{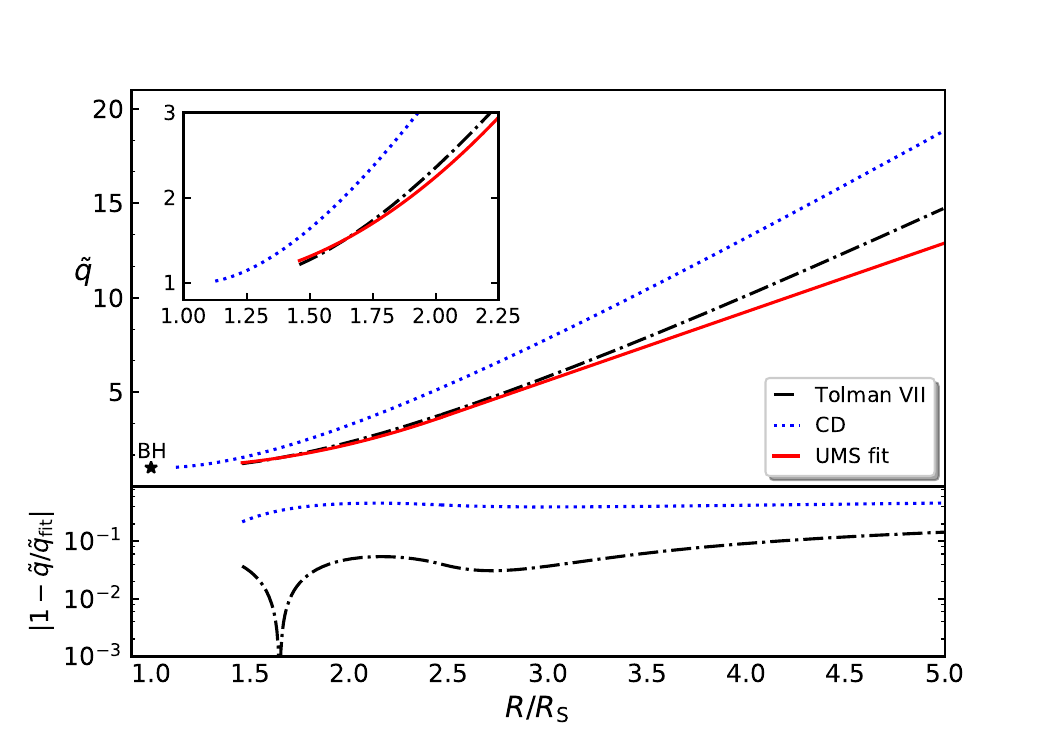}
\caption{The ‘Kerr factor' $\tilde{q}\equiv QM/J^2$, as a function of $R/\Rs$, for the T-VII solution (dashed black curve) and CD configurations (dotted blue curve). We also plot the analytic fitting formula for the selected EOSs considered by Urbanec et al. (UMS) \cite{Urbanec:2013fs} (solid red curve). The star indicates the Kerr BH value $\tilde{q}=1$. The bottom panel shows the relative error between the numerical data and the UMS fit.}
\label{fig:q}
\end{figure}

In~\fref{fig:q} we plot the ‘Kerr factor' $\tilde{q}\equiv QM/J^2$ \cite{Miller:1977}, as a function of the parameter $R/\Rs$, for the T-VII solution (solid black curve) and CD stars (dotted blue curve). The Kerr factor is relevant because it gives an account of the deviations of the external Hartle-Thorne metric away from the Kerr spacetime ($\tilde{q}_{\rm{BH}}=1$). First of all, we confirmed with an excellent agreement the results for homogeneous masses reported by \cite{Chandra:1974}. We observe that as the compactness increases, $\tilde{q}$ decreases monotonically and it seems to approach the Kerr BH value as $\beta\to 1$. For instance, for the T-VII spheres, for $\beta=\beta_{\mathrm{stab}}$ we obtain $\tilde{q}=1.215$ (see~\tref{tab:2}). A similar behaviour was also found for a number of realistic EOSs for NSs, including polytropes \cite{Yagi:2013awa,Urbanec:2013fs}. Thus, these results support the universality of the approach of $\tilde{q}$ to the Kerr value as the compactness approaches 1/2. However, let us stress out that the strict BH limit ($\beta\to 1$) cannot be considered from this solution or any other realistic EOS for NSs, no matter what value of the central energy density is chosen. It is worth noting that the $\tilde{q}-\mathcal{C}$ relation predicted by the CD configuration is quite different from the one predicted by the T-VII solution. 

By systematically studying the quadrupole moment for various EOSs for realistic NSs, Urbanec et al. \cite{Urbanec:2013fs} found an approximate universal behaviour between $\tilde{q}$ and $\beta$, which is almost insensitive to the EOS. Furthermore, they proposed an analytical fitting formula to describe this universal behaviour, which takes the form

\begin{equation}\label{qcfit}
\tilde{q} = \Big\{\begin{array} {lr} 
 \alpha_{1}\beta + \alpha_0,\quad &\beta>\beta_0,\\
 \delta(\beta-1)^2 + 1,\quad &\beta\leq \beta_0,
\end{array}
\end{equation} 

\noindent where $\delta=-\alpha_{1}^2/[4(\alpha_{1}+\alpha_{0}-1)]$, with the fitted coefficients $\alpha_0=-5.3$ and $\alpha_1=3.64$, and $\beta_0=2(1-\alpha_0)/\alpha_1$ is the matching point.

In order to compare the $\tilde{q}-\mathcal{C}$ relation for the T-VII solution with the one for realistic EOSs for NSs, in~\fref{fig:q} we include the fitting formula \eref{qcfit} with the coefficients found by \cite{Urbanec:2013fs} (solid red line). In the bottom panel, we show the absolute fractional difference.  We observe that for values of $\beta<4.15$ deviations are $10\%$, at most, in agreement with the fractional differences for data of realistic NSs \cite{Yagi:2016bkt}. In the relevant range of compactness for realistic NSs, i.e. $1.63<\beta<2.9$, relative errors vary between $\sim 0.02\%$ and $\sim 5\%$, at most. These differences are a bit higher compared with those we found for the normalised moment of inertia $I/MR^2$ in the same compactness regime. On the other hand, we observe that the $\tilde{q}-\mathcal{C}$ relation for CD configurations is quite different from the approximate universal fit.

\begin{figure}%[h!]
\centering
\includegraphics[width=0.6\linewidth]{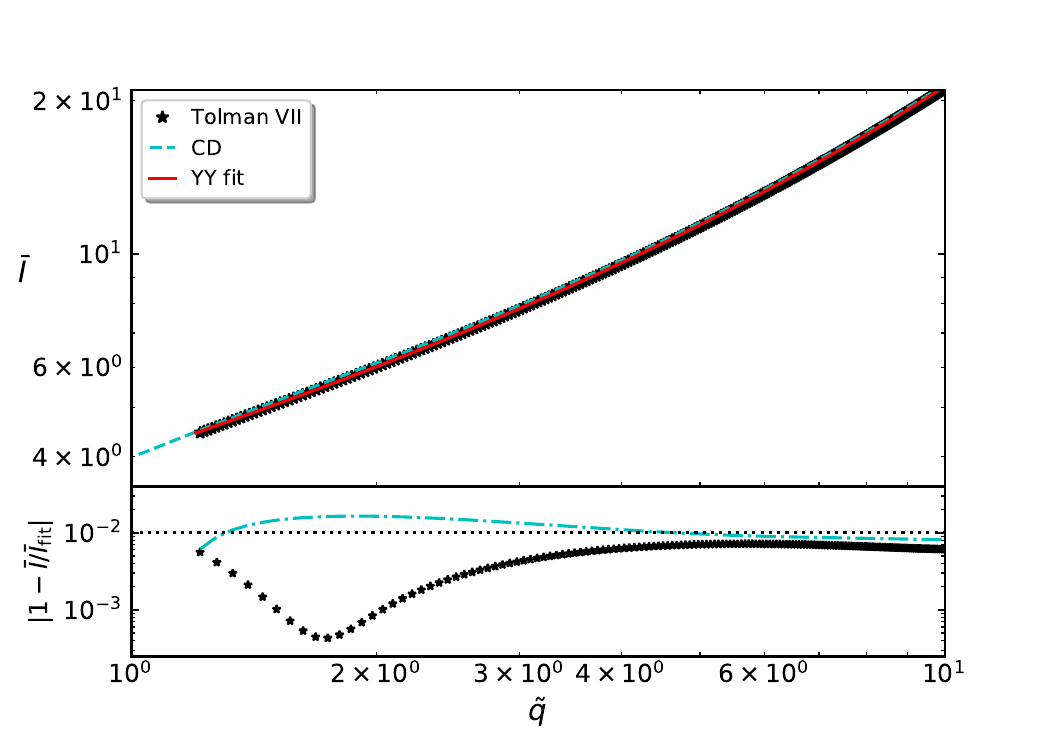}
\caption{$\bar{I}-\tilde{q}$ relation for the Tolman VII solution (solid black curve) and CD configurations (dotted cyan curve). We include the analytic fitting formula~\eref{iqfit} proposed by Yagi and Yunes \cite{Yagi:2016bkt} for realistic NSs (solid red curve). The star indicates the BH limit. The parameter being varied here is the central energy density, or equivalently the compactness, which increases as we move to the left of the panel. The bottom panel shows the fractional difference between the numerical results and the universal fit.}
\label{fig:iq}
\end{figure}

Finally, in~\fref{fig:iq} we present the $\bar{I}-\tilde{q}$ relation for the slowly rotating T-VII solution (starred black curve) and the CD configuration (dotted cyan curve). In order to compare with the $\bar{I}-\tilde{q}$ relation for realistic NSs, we plot the universal fit proposed by Yagi and Yunes \cite{Yagi:2016bkt} given by

\begin{equation}\label{iqfit}
\ln \bar{I} = a + b\ln\tilde{q} + c(\ln\tilde{q})^2 + d(\ln\tilde{q})^3 + e(\ln\tilde{q})^4,
\end{equation}

\noindent with the fitting coefficients: $a=1.393$, $b=0.5471$, $c=0.03028$, $d=0.01926$ and $e=4.434\times10^{-4}$. We observe that as the compactness of the star increases, (toward the left of the panel), the respective $\tilde{I}-\tilde{q}$ relations approach the BH limit (indicated by the star). In the bottom panel, we show the fractional difference between the corresponding numerical results and the fit. We observe that deviations of the $\bar{I}-\tilde{q}$ relation for the T-VII solution, and CD stars, are within $\mathcal{O}(1)\%$; thus, we conclude that, at the second order in $\Omega$, these analytical stellar models preserve the universality of the $\tilde{I}-\tilde{q}$ relation. 

\section{Concluding remarks}\label{sec:Conclud}

In this paper, we have presented an extension to slow rotation of the well-known Tolman VII (T-VII) solution \cite{Tolman:1939}. For this purpose, we employed the Hartle-Thorne perturbative framework at second order in the angular velocity $\Omega$ \cite{Hartle:1967he,Hartle:1968si}. In this approximation, the fractional rotational perturbations are fully determined by the parameter $\beta\equiv R/\Rs$, and the ratio $J/M^2$. Thus, we integrated the Hartle-Thorne (HT) equations for various values of the parameter $R/\Rs$, up to the limiting value for dynamical stability, and then determined the corresponding surface and integral properties for each model. As a test code, we solved the HT system for configurations with constant density (CD); we confirmed with excellent agreement the results reported by \cite{Chandra:1974,Reina:2015jia}. 

Let us summarize our main results. We determined the dimensionless moment of inertia $\bar{I}\equiv I/M^3$ for the T-VII model; our results are in excellent agreement with the analytical solution found by \cite{Jiang:2020uvb}. We found that, as the compactness of the configuration increases, $\bar{I}$ decreases monotonically and seems to approach the BH value $\bar{I}=4$, as $\beta$ approaches the limiting value for stability. We must stress, however, that the strict BH limit, i.e. $\beta\to 1$, cannot be taken from this solution, or any other realistic EOS for NS, no matter what value of the central energy density is chosen. We also found that the $\bar{I}-\mathcal{C}$ relation, for the T-VII solution, deviates up to $\mathcal{O}(10)\%$, with respect to the universal fit~\eref{YYfit}, in agreement with previous results for realistic EOSs for NSs \cite{Breu:2016ufb}. Furthermore, we found that the normalized moment of inertia $\tilde{I}\equiv I/MR^2$, within the relevant regime of compactness for realistic NSs, is in excellent agreement with the numerical results for realistic EOSs of NSs. Thus, in these respects, the T-VII solution provides a satisfactory approximate description of the interior of a NS.
 
We determined the ellipticity of the bounding surface for the T-VII spheroids, for fixed mass and angular momentum, and we found a reversal in its behaviour when $\beta\sim 2.65$. This seemingly counterintuitive effect, which is not predicted in Newtonian theory, was also found in homogeneous stars \cite{Chandra:1974} and polytropes \cite{Miller:1977}, in slow rotation. As it was argued by \cite{Abramowicz:1990a, Gupta:1996zva}, the reversal in the behaviour of the ellipticity is a consequence of the change in the behaviour of the rotational effects in the presence of strong gravitational fields, for instance, the reversal of the centrifugal force, which occurs even in the absence of dragging effects \cite{Abramowicz:1990b,Abramowicz:1990c,Abramowicz:1993irb}.   

We determined the Kerr factor $\tilde{q}$ for the T-VII spheroids and found that it decreases monotonically and seems to approach the Kerr value $\tilde{q}=1$, as the compactness tends to the BH limit. We compared our results with the universal  $\tilde{q}-\mathcal{C}$ relation~\eref{qcfit} for realistic NSs, and we found that relative errors are within $\mathcal{O}(10)\%$ which is consistent with the fractional differences of numerical solutions for realistic EOSs for NSs \cite{Yagi:2016bkt}. Finally, we determined the $\bar{I}-\tilde{q}$ relation for the slowly rotating T-VII solution, and we found that deviations, with respect to the universal fit~\eref{iqfit}, are within $\mathcal{O}(1)\%$. Thus, the T-VII solution preserves the universality of the $\bar{I}-\tilde{q}$ relation. It is worthwhile to remark that despite its simplicity,  the T-VII solution seems to approximate rather well the interior of realistic NSs, which are expected to have an intricate internal composition, characterised by different layers endowed with gradients of density and pressure. 

An alternative to improve the agreement between the T-VII solution and the universal fitting relations for realistic NSs, is to find some suitable modification to the T-VII model. Along this direction, \cite{Jiang:2019vmf} proposed a modified T-VII solution (MT-VII), by including an extra parameter $\alpha$ that allows the energy density to become a quartic function of $r$. These authors showed that this model provides a better agreement, for metric, pressure, and energy density, with the numerical results of realistic EOSs for NSs. However, in \cite{Posada:2022lij} we pointed out certain flaws of the MT-VII model, particularly the prediction of unphysical regions of negative pressure near the stellar surface, for certain values of the parameter $\alpha$ and compactness $\com$. Thus, inspired by the energy density profile introduced by \cite{Jiang:2019vmf}, we proposed the ‘exact modified Tolman VII' (EMT-VII) solution \cite{Posada:2022lij}, which alleviates the drawbacks of the original MT-VII model. Therefore, a further study would be to consider the EMT-VII model in slow rotation, determine its corresponding normalized moment of inertia and quadrupole moments, and then examine how these compare with the results for realistic EOSs for NSs. It is likely that for certain values of the parameter, $\alpha$ the agreement for $I$ and $\tilde{q}$ could improve significantly.\\ 

\ack
We thank John C. Miller for useful discussions and comments. We also thank the anonymous referees for providing valuable suggestions to improve this manuscript. The authors acknowledge the support of the Institute of Physics and its Research Centre for Theoretical Physics and Astrophysics, at the Silesian University in Opava. 

%\clearpage

\appendix
\section{Integral and surface properties of slowly rotating Tolman VII spheroids}
\label{appendix}
\setcounter{section}{1}

\begin{table}[h!]
\caption{\label{tab:2}Surface properties of slowly rotating Tolman VII spheroids, for some selective values of the parameter $\beta\equiv R/\Rs$. The main results are displayed in the accompanying figures. We employ the same units introduced by \cite{Chandra:1974,Miller:1977}. Here we include the corresponding factors $c$ and $G$ in case one wishes to recover the physical parameters. The stellar radius is measured in units of the Schwarzschild radius $\Rs\equiv 2GM/c^2$. The moment of inertia $I$ is measured in units of $M^3$; the normalized moment of inertia $I/MR^2$ is dimensionless; the change in mass $\delta M/M$ is measured in the unit $(GJ/M^2 c^3)^2$; the ‘Kerr factor' $\tilde{q}\equiv QM/J^2$ is dimensionless; the surface angular velocity $\varpi_{1}$, relative to the local inertial frame, is measured in the unit $GJ/M^3c^2$; the ellipticity $\varepsilon$ is measured in units of $(GJ/M^2 c^3)^2$.}
\begin{indented}
\item[]\begin{tabular}{@{}lllllll}
\br
$R/\Rs$ & $I/M^3$ & $I/MR^2$ & $\delta M/M$ & $\tilde{q}$ & $\varpi_{1}$ & $\varepsilon$ \\
\mr
  1.459 & 4.4619 & 0.5240 & 0.2098 & 1.2157 & 1.1489 & 0.5397 \\
  1.492 & 4.5598 & 0.5120 & 0.2223 & 1.2609 & 1.1522 & 0.5635 \\
  1.780 & 5.6323 & 0.4444 & 0.3160 & 1.8000 & 1.0657 & 0.7220 \\
  2.104 & 7.1747 & 0.4051 & 0.3852 & 2.6523 & 0.9002 & 0.8135 \\
  2.763 & 11.164 & 0.3656 & 0.4438 & 4.9041 & 0.6217 & 0.8539 \\
  3.690 & 18.537 & 0.3403 & 0.4440 & 8.7290 & 0.3917 & 0.7988 \\
  5.000 & 32.354 & 0.3235 & 0.4021 & 14.758 & 0.2312 & 0.6896 \\
\br
\end{tabular}
\end{indented}
\end{table}

\newpage
%%%%%%%%%%%%%%%%%%%% REFERENCES %%%%%%%%%%%%%%%%%%
\section*{References}
\bibliographystyle{iopart-num.bst}
\bibliography{references}
\end{document}